\documentclass[nonacm]{acmart}

\AtBeginDocument{%
  }

\newtheorem{dfn}{Definition}

\usepackage{multirow}
\usepackage{algorithmic}
\usepackage{algorithm}
\usepackage{listings}
\usepackage{longtable}
\usepackage{tabularray}
\usepackage{multirow}
\usepackage{tabularx}

\title{A novel DFS/BFS approach towards link prediction} 

\author{Jens Dörpinghaus}
\email{doerpinghaus@uni-koblenz.de}
\orcid{0000-0003-0245-7752}
\affiliation{%
  \institution{Federal Institute for Vocational Education and Training}
  \city{Bonn}
  \country{Germany}
}
\affiliation{%
  \institution{University of Koblenz}
  \city{Koblenz}
  \country{Germany}
}
\author{Tobias Hübenthal}
\affiliation{%
  \institution{WTW}
  \city{Cologne}
  \country{Germany}
}
\author{Denis Stepanov}
\affiliation{%
  \institution{Department of Mathematics and Computer Science, University of Cologne}
  \city{Cologne}
  \country{Germany}
}

\begin{document}

\renewcommand{\shortauthors}{Dörpinghaus et al.}
\begin{abstract}
Knowledge graphs have been shown to play a significant role in current knowledge mining fields, including life sciences, bioinformatics, computational social sciences, and social network analysis. The problem of link prediction bears many applications and has been extensively studied. However, most methods are restricted to dimension reduction, probabilistic model, or similarity-based approaches and are inherently biased.  In this paper, we provide a definition of graph prediction for link prediction and outline related work to support our novel approach, which integrates centrality measures with classical machine learning methods. We examine our experimental results in detail and identify areas for potential further research.  Our method shows promise, particularly when utilizing randomly selected nodes and degree centrality.
\end{abstract}

\maketitle
\section{Introduction}

Knowledge graphs play a crucial role in contemporary knowledge mining settings, including life sciences, bioinformatics, computational social sciences, and social network analysis. 
Contextual information is critical for modeling complicated systems like the labor market and is also prevalent in NLP and knowledge discovery tasks, as it significantly affects the precise meaning of expressions and data queries. Here we present results on link prediction in knowledge graphs that utilize contextual information, such as the graph's topology.

Link prediction involves methods that estimate edge existence between two nodes in a graph and is widely used in research.Graph structures, particularly knowledge graphs, provide several advantages for targeted re-extraction and knowledge integration.  Biomedical literature and text mining are utilized to construct knowledge graphs,  see for example  \cite{dorpinghaus2018scaiview,dorpinghaus2019knowledge,dorpinghaus2019semantic}.Link prediction is significant in life sciences, medical research, and their related disciplines, as it enhances formal interconnectedness. Link prediction is employed in social networks to forecast social interaction \cite{daud2020applications,zhao2022novel} and is applied more broadly across the social sciences \cite{yang2015evaluating,dorpinghaus2023rule,dorpinghaus2021soziale} and in particular in labor market research \cite{fischer2024web} or modeling longitudinal data \cite{dorpinghaus2024towards}.

There are several methods for link prediction. While dimension reduction based methods show promising results, they rely on a method that cannot be applied universally without bias. Probabilistic model-based methods do not rely solely on graph structure. Similarity-based methods can also rely on additional information, but do not generalise. Therefore, we will now introduce a generalised method that allows experiments with learning graph structures including similarity approaches.
However, previous research has shown that the performance on some subgraphs is very poor and the time complexity is high. In order to identify limitations and understand the underlying problems, we will investigate these issues.

As such, our research question is: How can a combination of depth and breadth search methods be employed for the purpose of identifying node neighborhood structure, or context, and ultimately enhance machine learning techniques utilized for edge prediction within graphs? This approach raises several sub-research questions. For example, how does the performance compare to other state-of-the-art methods? What is the impact of node selection in the neighborhood on the results of this method? How many neighbors must be considered, and what is the impact of this?

The paper is divided into six sections. After an introduction, the second section describes the problem statement. The third section gives a brief overview of the state of the art, related work and background for our novel approach. The fourth section describes the novel link prediction approach, with the experimental results on both artificial and real scenarios in the following section.
Our conclusions and outlook are given in the last sections. 

\section{Problem statement}
We restrict ourselves to the case where $G=(V,E)$ is an unweighted, undirected graph. We want to examine $G$ and, given a scoring function $\phi:(V',E')\rightarrow [0,1]$, we can estimate the plausibility of any subgraph $\hat{G} \subseteq G$, i.e. $\phi$ returns a higher score, if $\hat{G}$ exists. This includes, for example, the existence of one (or more) new links as well as the existence of new vertices that complete the main graph $G$.

Let $\mathcal{L}_{\phi,\epsilon}$ be the function that predicts and adds new graph structure to $G$. This implies, given a subgraph $ \hat{G}=(\hat{V}, \hat{E}) \subseteq G$ with $ \hat{V}\subseteq V,\hat{E} \subseteq E$,

\[\mathcal{L}(\hat{G}) = G\bigcup_{X\subseteq K(\hat{G}),\, \phi(X)\geq \epsilon} X \]
with a threshold  $\epsilon \in [0,1]$. Here, $K(\hat{G})$ refers to the complete graph on $V(\hat{G})$. Note that it is not necessary to set $X\subseteq K(\hat{G})$, and that $\phi$ could also be used to predict new nodes. However, here, we restrict the results to edges.

For nodes, this problem basically refers to the well-known definition of link prediction:
\[\mathcal{L}(e) = \begin{cases}
G & \phi (e)<\epsilon\\
G \cup e & \phi (e)\geq\epsilon\\
\end{cases} \]

\begin{dfn}[Graph Prediction]
	Let $G=(V, E) $ be a graph, where $V $ is the set of vertices and $E \subseteq V \times V $ is the set of existing links. Let $\mathcal{L}_{\phi,\epsilon}$ be a prediction function. We define $\mathcal{L}(G) =G' $ with $G \subseteq G'$ and $G'$ as the graph with new links.
\end{dfn}

This general definition could be specified and modified according to the task and objectives. We will now introduce two methods to predict a new structure of $G$ utilising previously predicted edges. This, however, needs functions $\phi$ that rely on graph structures.
\begin{itemize}
	\item \textbf{Iterative methods:}
	After a new link has been found in $\hat{G} $ and added to $G$, the state of $G $ at that time is stored and used in the next iteration of the method.
	Let $ t=\{t_0, \dots , t_n\} $ be the set of times and $ G(t_i) $ be the state of $ G $ at time $ t_i $ for $ i \in \{0, \dots , n\} $:
	\begin{equation*}
		G(t_{i+1})=\mathcal{L}(G(t_i)))
	\end{equation*}
	\item \textbf{Non-iterative methods:}
	In this case of the prediction process, $G$ is iterated until there are no more new links  to predict and a final state of $G$ is the output.
	\begin{equation*}
		G_{end}=\mathcal{L}(G_{start})
	\end{equation*}
\end{itemize}

Table \ref{tab0:illu} shows examples of the function $\mathcal{L}$, the corresponding structure of $\hat{G}$ and $\hat{G}'$ and practical relevance. Illustration of two functions $ \mathcal{L}_1 $ and $ \mathcal{L}_2 $ which predict different subgraphs:  $ \mathcal{L}_1 $ is a generic prediction function for nodes and edges, while $ \mathcal{L}_2 $ predicts edges, limiting to one edge refers to the classical definition of link prediction.
\begin{table}[t]
\caption{Illustration of two functions $ \mathcal{L}_1 $ and $ \mathcal{L}_2 $ which predict different subgraphs:  $ \mathcal{L}_1 $ is a generic prediction function for nodes and edges, while $ \mathcal{L}_2 $ predicts edges, limiting to one edge refers to the classical definition of link prediction. }\label{tab0:illu}
	\centering
\begin{tabular}{|l|c|c|}
\hline
 $ \mathcal{L} $   & $ \hat{G}' $ & $ \hat{G} $ \tabularnewline
\hline
\hline
$ \mathcal{L}_1 $ & \includegraphics[width=0.08\textwidth]{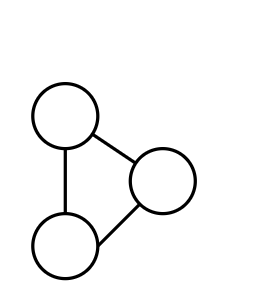}        & \includegraphics[width=0.08\textwidth]{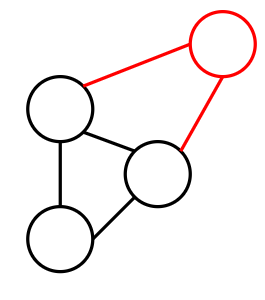} \tabularnewline
\hline
\multirow{2}{*}{$ \mathcal{L}_2 $} &\includegraphics[width=0.05\textwidth]{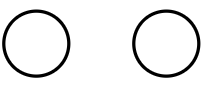}        &  \includegraphics[width=0.05\textwidth]{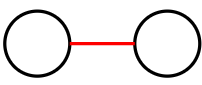}\tabularnewline
\cline{2-3} \cline{3-3}
 & \includegraphics[width=0.1\textwidth]{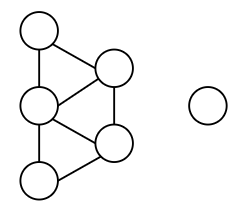}        & \includegraphics[width=0.1\textwidth]{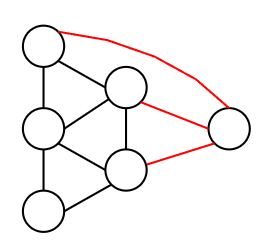}\tabularnewline
\hline
\end{tabular}

\end{table}

Link prediction is a subproblem of complex prediction of another structure in $G$. Namely, it is restricted to the prediction of links in a given graph. In the following chapter the problem is considered in more detail. The methods that can be used to solve this problem are presented and then applied to a practical task.

\section{Related work}

The variety of different methods to solve this problem is classified in different ways in the literature. We will follow \cite{9} and \cite{rossi2021knowledge} for classification of related work, see Table \ref{tab00:classes}. Widely discussed are methods based on graph embeddings, which are low-dimensional representations of the graph. Numerous novel models, approaches and architectures were proposed in the last years.
For graph embeddings, Rossi et al. mention tensor and matrix factorisation, geometric and deep learning approaches.
We need to add similarity-based methods and probabilistic model-based methods. However, other approaches like learning or clustering-based also exist but will not be considered in this work \cite{hasan2011survey,huang2010link}.

\begin{table*}[t]
\footnotesize 
\caption{Classes of Link Prediction approaches.}\label{tab00:classes}
\centering
\begin{tabular}{|c|c|c|c|}
\hline
Dimensionally reduction-based & Similarity-based & Probabilistic model-based methods & Other approaches\tabularnewline
\hline
\hline
Tensor and matrix factorisation & Local similarity & Relational Bayesian Networks & Learning-based\tabularnewline
\hline
Geometric approaches & Global similarity & Relational Markov Networks & Clustering-based\tabularnewline
\hline
Deep learning approaches & ... & ... & ...\tabularnewline
\hline
... &  &  & \tabularnewline
\hline
\end{tabular}
\end{table*}

\subsection{Dimensionally reduction-based}
Graph embeddings are widely used for link prediction, e.g. in knowledge graphs \cite{rossi2021knowledge,wang2021survey,crichton2018neural}. In general, they use a dimensionality reduction technique: $D$-dimensional nodes in the graphs are mapped to a lower $d$-dimensional mapping space ($d \ll D$) by preserving the neighbourhood structures of nodes so that similar nodes (in the original network) have a similar representation in the mapping space. Rossi et al. provide an extensive overview of these methods, see \cite{rossi2021knowledge}.

Methods in the class \emph{tensor and matrix factorisation} represent the graph and connections between nodes in the form of a 3D adjacency matrix and factorise this matrix to obtain the embedding of nodes and edges, see \cite{acar2009link,rossi2021knowledge}. Adjacency matrices, Laplace matrices, transition matrices and others are used to represent the links, depending on the specific method. The approaches to factorising the representative matrix vary depending on the properties of the matrix, for example bilinear or non-linear models. For example, if the resulting matrix is positive semidefinite, such as the Laplace matrix, eigenvalue decomposition can be used  \cite{5,dunlavy2011temporal}.

Yet another approach is known as DeepWalk \cite{cai2018novel,berahmand2021modified,13}.
This method is based on random walks in the graph. A random walk is a stochastic process with random variables $X_{v_i}^1,X_{v_i}^2, \dots ,X_{v_i}^k$ such that $X_{v_i}^k$ is a randomly chosen node from the neighbourhood set of $v_{k-1}$. The similarity function $S(v,u)$ is defined as the probability $p(v|u)$ of reaching a node $v$ in a random walk starting at node $v_0=v$, see \cite{14}.
A special case is Node2vec, see \cite{6}. The graph structure in this method is captured with a 2nd degree random walk (or a 2nd degree Markov chain). Other scholars use HMM \cite{fechner2024ensemble}.

\subsection{Similarity-based}
Similarity-based methods are the easiest to implement. For each pair of nodes $v,u$ a similarity measure $S(v,u)$ is computed. $S(v,u)$ is determined from the structural or node properties of the node pair $(v,u)$. The predicted links are assigned scores according to similarity measures between $v$ and $u$. The connection with the higher score represents the predicted connection $(v,u)$. The similarity measures between each pair can be calculated using several properties of the network, one of which is structural.\cite{9}
We group these methods into the following categories: Local and global similarity and centrality measures.

Local similarity measures are generally computed using information about common immediate neighbours and node degree. Examples are Common Neighbours \cite{12},  Jaccard Coefficient \cite{9} and Adamic/Adar Index \cite{1}.  In contrast, global similarity measures are computed taking into account all the topological information of a network. Examples are Katz Index\cite{8} or SimRank\cite{7}.

Centrality measures are used in graph analysis to describe the relative influence of a node on the network \cite{3}.  These measures cannot predict a link by themselves and can be used as additional features together with similarity measures. Some scholars investigate the influence of centrality and similarity measures as well as other features in predicting links using machine learning. The following centrality measures are considered:
\begin{enumerate}
	\item Degree centrality:
	Degree centrality is defined as the number of connections of each node in a graph.
	\item Betweenness centrality:
	This measure indicates how many shortest paths between other nodes pass through a node. Nodes with high betweenness centrality tend to connect otherwise unconnected subsets of a graph. 
	\item Closeness centrality:
	The closeness centrality of a node measures its average distance to all other nodes. Nodes with a high measure have the shortest distances to all other nodes. 
\end{enumerate}

\subsection{Probabilistic model-based methods}

Here, probabilistic models are used to optimise objective functions. However, they rely on additional data, for example node and edge attributes, see \cite{9}. Thus, they cannot rely on structural information only. Several probabilistic models \cite{wang2007local,yu2006stochastic} and maximum likelihood models
\cite{clauset2008hierarchical} were proposed.

A special case was introduced in \cite{dorpinghaus2022novel}: Here, Conditional Random Fields, a special case of Markov Random Fields, were used to build a model to predict links. Nodes were selected using centrality measures, which are a similarity-based approach. According to our knowledge, no more work was carried out combining probabilistic and similarity-based methods. 

To summarise, while dimensionally reduction-based methods show promising results, they rely on a method which cannot be applied generally without a bias. Probabilistic model-based methods do not rely solely on graph structures. However, similarity-based methods may also rely on additional information but do not generalise. Thus, we will now introduce a generalised method which allows experiments with learning of graph structures including similarity approaches.



\section{Method}

 When constructing a graph-based learning model for link prediction, the selection of accurate features is of utmost significance. These chosen features must aptly characterize the graph's topology. The issue of identifying the appropriate features for link prediction has been discussed earlier.  In the following, a method will be presented that randomly picks parts of the topology or uses similarity approaches to solve this problem.  

\begin{figure}[t]
	\centering
	\includegraphics[scale=0.4]{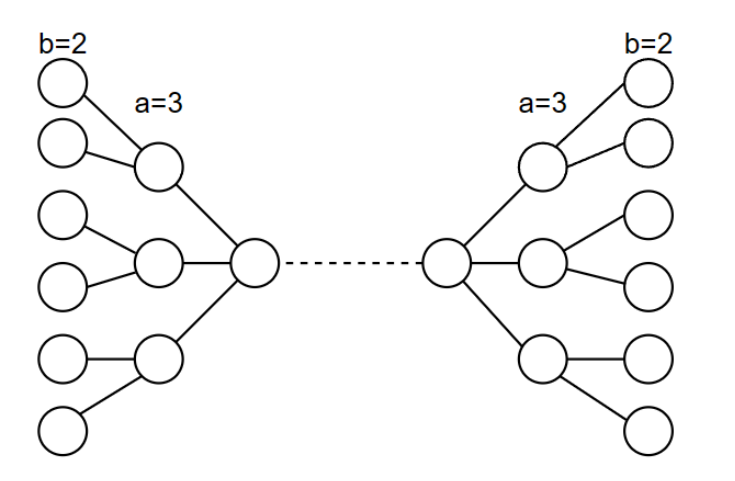}
	\caption{In illustration of the proposed approach: Given two nodes $u,v\in V$ and $a=3$ and $b=2$. For both $u,v$ we consider $a$ neighbours, and for the depths of $b$, for each neighbour again $a$ neighbours.}
	\label{model_abb}
\end{figure}

Considering a graph $G=(V,E)$ and a pair of nodes $u$ and $v\in V$ in the graph, a set will be defined for each node consisting of $a$ neighbors at distance $b$.  These sets will be utilized as features for machine learning models, in which the choice of $a$ and $b$ determines whether the model represents a depth-first or breadth-first search. Figure \ref{model_abb} presents an illustration of a pair of nodes and the parameters $a=3$ and $b=2$. 

Of course, it is not always possible to compute unique sets of neighbors for every pair of nodes in the graph and for every parameter $a$ and $b$. Our model enables first-degree neighbors to appear repeatedly as features, meaning that important information  concerning relationships to $b$-degree neighbors won't be lost in situations where we lack enough unique, but do possess some shared degree neighbors (like for large $a$, for instance), for a given pair of nodes, $u$ and $v$.

When selecting neighbors with a higher degree, the code verifies if the node got visited before. In case there are no more available neighbors and $k$ neighbors have already been visited for $a$ or $b$, the neighbor set will get completed using $a-k$ or $b-k$ zeros. When all the neighbors' spaces are already visited or empty, the neighbor set size will be equal to $a$ or $b$ with zeros within it.

It is important to consider the criteria used to select neighbors. In our study, we analyze two potential methods:
\begin{itemize}
	\item \emph{Selecting neighbors randomly}. Neighbors are selected randomly for each node, and the first $a$ or $b$ elements of each set are utilized as features in the training set.
	\item The neighbor set is sorted, and the $a$ \emph{nodes with the highest rankings are selected}. The nodes in this set are sorted in descending order based on their centrality measures, specifically considering closeness and betweenness centralities. The first $a$ or $b$ items from the sorted set are added to the training set. 
\end{itemize}
The next step is to construct a machine learning pipeline, which involves creating a feature vector, denoted as $X$, and a target vector, denoted as $y$.  The data is then partitioned into four subsets: $Xtrain$ and $ytrain$ for training, and $Xtest$ and $ytest$ for testing. Next, a procedure is trained on the training sets $Xtrain$ and $ytrain$ and used to predict outcomes for the test set $Xtest$. The predicted results vector, $y_{pred}$, will be compared to the set $ytest$ when evaluating the model in the following chapter.

\begin{algorithm}[t]
\begin{algorithmic}[1]
\caption{\textsc{Create} $a,b$ nodeset for $u,v\in V$} \label{alg:create}
\REQUIRE Graph $G=(V,E)$ and $u,v\in V$, $a\in\mathbb{N}^+$, $b\in\mathbb{N}$
\ENSURE $x,\tilde{y}$
\STATE $y=\begin{cases}
0 & (u,v)\not\in E \\
1  & (u,v)\in E\\
\end{cases}$
\STATE $N\leftarrow oN(u)[:a] $ 
\FOR{$i\in \{1,b\}$}
\FOR{$j\in \{1,a\}$}
\STATE $N\leftarrow oN(N[(i\cdot a)+j])[:a] $
\ENDFOR
\ENDFOR
\STATE $M\leftarrow oN(v)[:a] $
\FOR{$i\in \{1,b\}$}
\FOR{$j\in \{1,a\}$}
\STATE $M\leftarrow oN(M[(i\cdot a)+j][:a] $
\ENDFOR
\ENDFOR
\RETURN $x=N+M+u+v$, $\tilde{y}=y$
\end{algorithmic}
\end{algorithm}

\begin{algorithm}[t]
\begin{algorithmic}[1]
\caption{\textsc{Create} $X,y$ representation for $G$} \label{alg:create2}
\REQUIRE Graph $G=(V,E)$, $a\in\mathbb{N}^+$, $b\in\mathbb{N}$
\ENSURE $X, y$
\STATE $X\leftarrow []$, $y\leftarrow []$
\FOR{$i\in (0,|V|)$}
\FOR{$j\in (0,i)$}
\STATE $u=V[i]$, $v=V[j]$
\IF{$u\neq v$}
\STATE $x,\tilde{y}=$\textsc{Create}$(G,u,v,a,b)$
\STATE $X\leftarrow x$, $y\leftarrow \tilde{y}$
\ENDIF
\ENDFOR
\ENDFOR
\RETURN $X,y$
\end{algorithmic}
\end{algorithm}

To generate $X$ and $y$, Algorithm \ref{alg:create2} is presented. For each combination of nodes in $G$, $X$ and $y$ are filled using the breadth-depth approach described above. See Algorithm \ref{alg:create} for pseudocode.  Generally, the runtime is cubic in $n=|V|$, as the runtime of limited breadth and depth search is linear in $n$. We repeat this for all combinations of nodes, which leads to a worst-case scenario of $\mathcal{O}(n^3)$.To optimize the process, each pair is considered only once. Thus, only the upper diagonal matrix is traversed.

\section{Experimental results}

We utilized Python to implement our software and evaluated the underlying approach using Jupyter Notebooks. Additionally, we employed various libraries such as NetworkX, Scikit-learn, Matplotlib, NumPy, and Pandas for embedding establishment.

To assess the performance of our model, we compared it to the Node2Vec method.  As previously discussed, embeddings are expected to furnish more precise predictions in comparison to other models.For this purpose, we utilize the node2vec library, located at \url{https://github.com/eliorc/node2vec} and referenced in \cite{6}. 

The most important measures for evaluation are recall and precision, defined as follows:
\[	Recall=\frac{TP}{FN+TP}\]
\[	Precision=\frac{TP}{FP+TP}\]
Here \textit{TP} means true positive results, \textit{FN} false negatives, and \textit{FP} false positives. Therefore, $FP+TP$ encompasses all positive results, and $FP+TP$ are all samples that should have been identified as positive.
$Recall$  indicates the proportion of correctly classified data, while $Precision$ indicates the proportion of data classified as positive that is actually positive.

The $F_1$-score (also balanced $F$-score or $F$-measure) calculates the harmonic mean of $Precision$ and $Recall$.The optimal value is 1, while a score of 0 indicates poor performance. The formula for calculating $F$-measure is:  

	\[F_1 = 2 \cdot \frac{\textit{Precision} \cdot \textit{Recall}}{\textit{Precision} + \textit{Recall}},\]

For each graph described in the next subsection, we computed four experiments:
\begin{itemize}
\item An $a,b$-model with selection of neighbours by betweenness centrality,
\item by degree centrality,
\item with randomly chosen neighbours.
\item A state-of-the art  embeddings approach using Node2Vec.
\end{itemize}
We limit our experiments for $a,b$-model to a maximum of $a=5$ and $b=5$ for computational reasons.

\subsection{Data}

For testing purposes, we utilized a small in-house knowledge graph (Graph 1) and two subsets of a Facebook social network, consisting of circles and ego networks (Graphs 2+3) as presented in \cite{leskovec2012learning}.  
Table \ref{tab:2} displays the primary parameters of the graphs including the number of nodes and edges, the average node degree, and the source. 
\begin{table}[t]
	\centering
    \caption{Some characteristics of the networks used to evaluate our approach. }\label{tab:2}
	\begin{tabular}{|l|rrrl|}
\hline
		& Nodes & Edges & Average node degree & Source \\
        \hline
		Graph 1 &        151       &       235        &         3.11                &   in-house KG  \\
		Graph 2 &        61       &          270     &           8.85             &  \cite{leskovec2012learning}     \\
		Graph 3 &        333      &         2519      &            15.13             &     \cite{leskovec2012learning}  \\
	\hline
    \end{tabular}
\end{table}

Using graphs as data for learning can have a drawback, as graphs are typically sparse, leading to an imbalance.  For instance, let us take graph 1, composed of 151 nodes. The size of the adjacency matrix is 22,801 entries ($151 \times 151$). To learn, only the entries above the diagonal are utilized, resulting in 11,325 entries ($\frac{22,801-151}{2}$). Out of these, only 235 indicate a connection. We will examine how this impacts performance.

\begin{table}[t]
\centering
\caption{Results of \emph{Node2Vec} method for graphs 1-3.}\label{tab:emb}
\begin{tabular}{|l|r|r|r|}
\hline
 & Graph 1 & Graph 2 & Graph 3\tabularnewline
\hline
\hline
Precision & 1.00 & 0.89 & 0.89\tabularnewline
\hline
Recall & 0.85 & 0.69 & 0.37\tabularnewline
\hline
$F_{1}$-Score & 0.92 & 0.78 & 0.52\tabularnewline
\hline
\end{tabular}
\end{table}

\subsection{Results}

\begin{figure}[p]
\centering
\includegraphics[width=0.45\textwidth]{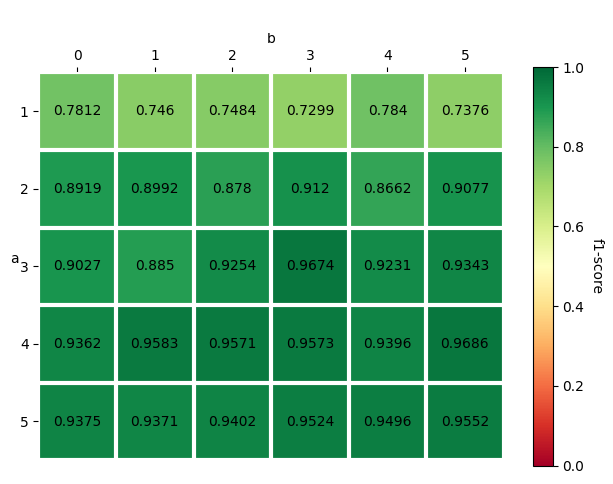}
\caption{$F_1$-score heatmap for $a,b\leq 5$ and graph 1 and betweenness centrality.}\label{img:resgr1a}

\vspace*{0.5cm}
\includegraphics[width=0.45\textwidth]{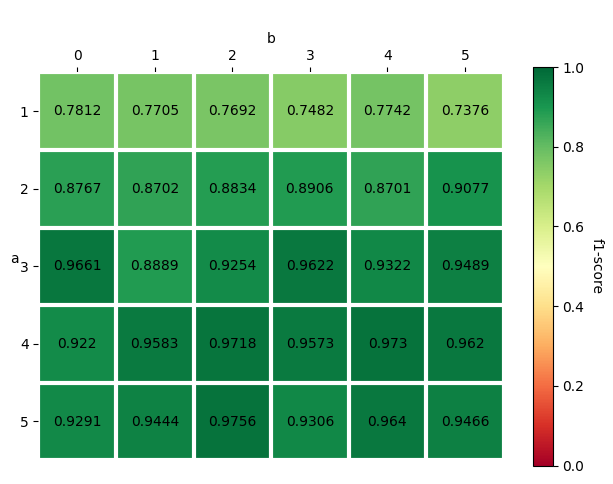}
\caption{$F_1$-score heatmap for $a,b\leq 5$ and graph 1 and degree centrality.}\label{img:resgr1b}

\vspace*{0.5cm}
\includegraphics[width=0.45\textwidth]{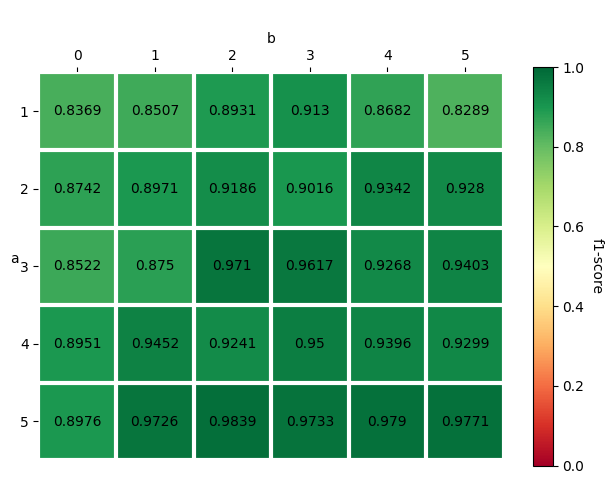}
\caption{$F_1$-score heatmap for $a,b\leq 5$ and graph 1 and randomly chosen nodes.}\label{img:resgr1c}
\end{figure}

\begin{figure}[p]
\centering
\includegraphics[width=0.45\textwidth]{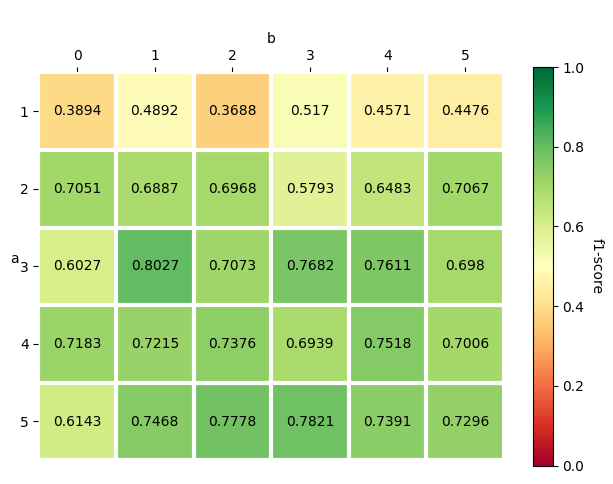}
\caption{$F_1$-score heatmap for $a,b\leq 5$ and graph 2 and betweenness centrality.}\label{img:resgr2a}

\vspace*{0.5cm}
\includegraphics[width=0.45\textwidth]{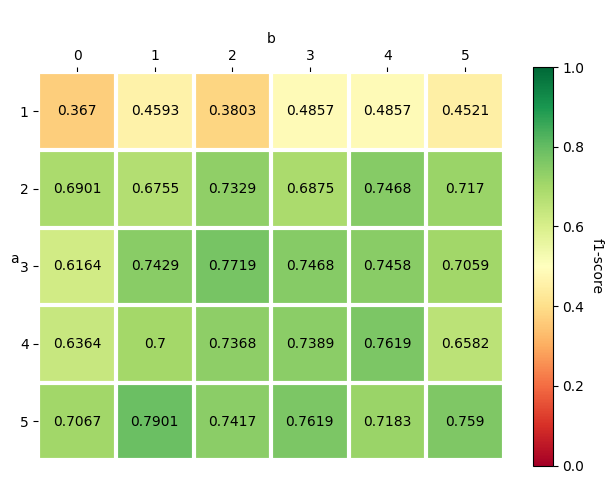}
\caption{$F_1$-score heatmap for $a,b\leq 5$ and graph 2 and degree centrality.}\label{img:resgr2b}

\vspace*{0.5cm}
\includegraphics[width=0.45\textwidth]{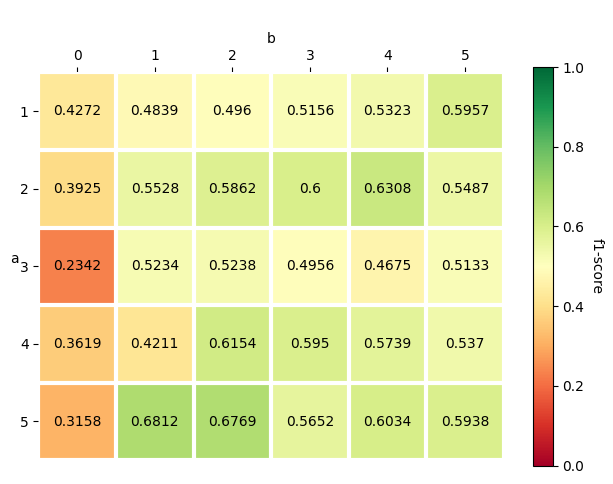}
\caption{$F_1$-score heatmap for $a,b\leq 5$ and graph 2 and randomly chosen nodes.}\label{img:resgr2c}
\end{figure}

\begin{figure}[p]
\centering
\includegraphics[width=0.45\textwidth]{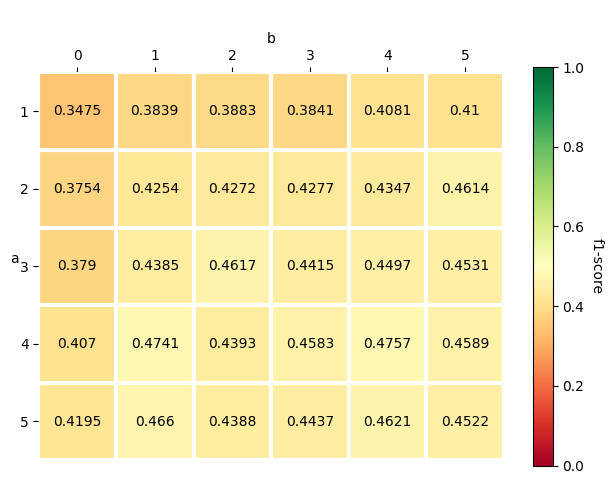}
\caption{$F_1$-score heatmap for $a,b\leq 5$ and graph 3 and betweenness centrality.}\label{img:resgr3a}
\vspace*{0.5cm}

\includegraphics[width=0.45\textwidth]{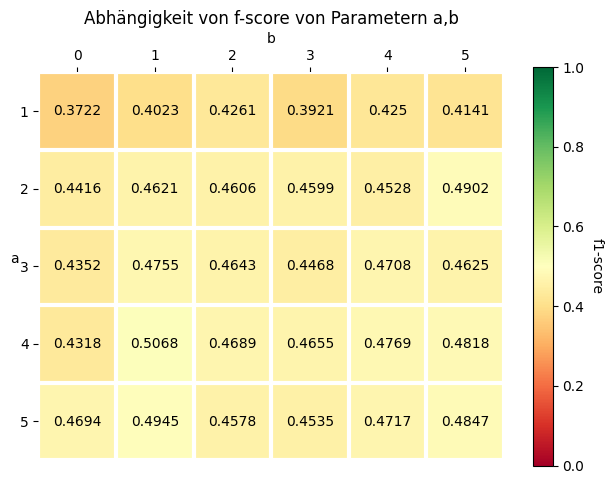}
\caption{$F_1$-score heatmap for $a,b\leq 5$ and graph 3 and degree centrality.}\label{img:resgr3b}

\vspace*{0.5cm}
\includegraphics[width=0.45\textwidth]{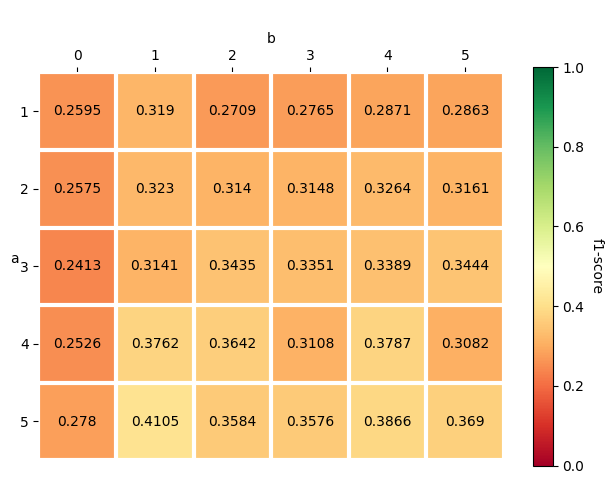}
\caption{$F_1$-score heatmap for $a,b\leq 5$ and graph 3 and randomly chosen nodes.}\label{img:resgr3c}
\end{figure}

First, we will discuss the results of the initial three approaches and compare the $F_1$-score as a heatmap for $a$ and $b$ values that are less than or equal to 5, across various graphs.  The results for graph 1 will be displayed in Figures \ref{img:resgr1a}-\ref{img:resgr1c}, graph 2 in Figures \ref{img:resgr2a}-\ref{img:resgr2c}, and graph 3 in Figures \ref{img:resgr3a}-\ref{img:resgr3c}.

\begin{table}[t]
\centering
\caption{Precision, Recall and $F_1$-score  for $a,b\leq 5$ and graph 1 and randomly chosen nodes.}\label{tab:gr1-rand}
	\begin{tabularx}{\linewidth}{|l|XXXXXX|}
    \hline
		\, & b=0                      & b=1                      & b=2                      & b=3                      & b=4                      & b=5                      \\
        \hline
        a=1                            & 0.8429\newline 0.831\newline0.8369  & 0.8636\newline0.8382\newline0.8507 & 0.9467\newline0.8452\newline0.8931 & 0.9545\newline0.875\newline0.913   & 0.918\newline0.8235\newline0.8682  & 0.875\newline0.7875\newline0.8289  \\
        \hline
        a=2                            & 0.9565\newline0.8049\newline0.8742 & 0.8841\newline0.9104\newline0.8971 & 0.9518\newline0.8876\newline0.9186 & 0.9322\newline0.873\newline0.9016  & 0.9467\newline0.9221\newline0.9342 & 0.9508\newline0.9062\newline0.928  \\
        \hline
        a=3                            & 0.8596\newline0.8448\newline0.8522 & 0.98\newline0.7903\newline0.875    & 0.9571\newline0.9853\newline0.971  & 0.9565\newline0.967\newline0.9617  & 0.9194\newline0.9344\newline0.9268 & 0.9692\newline0.913\newline0.9403  \\
        \hline
		   a=4     & 0.9143\newline0.8767\newline0.8951 & 0.9583\newline0.9324\newline0.9452 & 0.9178\newline0.9306\newline0.9241 & 0.95\newline0.95\newline0.95       & 0.9459\newline0.9333\newline0.9396 & 0.9241\newline0.9359\newline0.9299 \\
        \hline
        a=5                            & 0.8906\newline0.9048\newline0.8976 & 0.9726\newline0.9726\newline0.9726 & 0.9839\newline0.9839\newline0.9839 & 0.9865\newline0.9605\newline0.9733 & 0.9589\newline1.0\newline0.979     & 1.0\newline0.9552\newline0.9771 \\
        \hline

	\end{tabularx}

    \vspace*{0.6cm}
    \caption{Precision, Recall and $F_1$-score  for $a,b\leq 5$ and graph 2 and randomly chosen nodes.}\label{tab:gr2-rand}
    \begin{tabularx}{\linewidth}{|l|XXXXXX|}
    \hline
		\, & b=0                      & b=1                      & b=2                      & b=3                      & b=4                      & b=5                      \\
        \hline
    \hline
        a=1                            & 0.6286\newline0.3235\newline0.4272 & 0.6522\newline0.3846\newline0.4839 & 0.7381\newline0.3735\newline0.496  & 0.6346\newline0.4342\newline0.5156 & 0.6471\newline0.4521\newline0.5323 & 0.7368\newline0.5\newline0.5957    \\
        \hline
	 a=2                            & 0.84\newline0.2561\newline0.3925   & 0.8293\newline0.4146\newline0.5528 & 0.9714\newline0.4198\newline0.5862 & 0.9231\newline0.4444\newline0.6    & 0.8723\newline0.494\newline0.6308  & 0.7561\newline0.4306\newline0.5487 \\
     \hline
	 a=3                            & 0.8125\newline0.1368\newline0.2342 & 0.9032\newline0.3684\newline0.5234 & 0.9429\newline0.3626\newline0.5238 & 0.9032\newline0.3415\newline0.4956 & 0.9474\newline0.3103\newline0.4675 & 0.7838\newline0.3816\newline0.5133 \\
     \hline
 a=4                            & 0.7308\newline0.2405\newline0.3619 & 0.96\newline0.2697\newline0.4211   & 0.9\newline0.4675\newline0.6154    & 0.8571\newline0.4557\newline0.595  & 0.8919\newline0.4231\newline0.5739 & 0.9355\newline0.3766\newline0.537  \\
 \hline
 a=5                            & 0.6\newline0.2143\newline0.3158    & 0.9216\newline0.5402\newline0.6812 & 0.9565\newline0.5238\newline0.6769 & 0.9286\newline0.4062\newline0.5652 & 0.875\newline0.4605\newline0.6034  & 0.9048\newline0.4419\newline0.5938  \\
        \hline

	\end{tabularx}

    \vspace*{0.6cm}
    \caption{Precision, Recall and $F_1$-score  for $a,b\leq 5$ and graph 3 and randomly chosen nodes.}\label{tab:gr3-rand}
    \begin{tabularx}{\linewidth}{|l|XXXXXX|}
    \hline
		\, & b=0                      & b=1                      & b=2                      & b=3                      & b=4                      & b=5                      \\
        \hline
        a=1                            & 0.5018\newline0.175\newline0.2595  & 0.5661\newline0.2221\newline0.319  & 0.5551\newline0.1792\newline0.2709 & 0.5318\newline0.1868\newline0.2765 & 0.5515\newline0.194\newline0.2871  & 0.5458\newline0.194\newline0.2863  \\
        \hline
	 a=2                            & 0.5161\newline0.1716\newline0.2575 & 0.7536\newline0.2055\newline0.323  & 0.7512\newline0.1985\newline0.314  & 0.7171\newline0.2016\newline0.3148 & 0.7315\newline0.2101\newline0.3264 & 0.7887\newline0.1977\newline0.3161 \\
     \hline
	 a=3                            & 0.6529\newline0.148\newline0.2413  & 0.75\newline0.1986\newline0.3141   & 0.7778\newline0.2204\newline0.3435 & 0.803\newline0.2117\newline0.3351  & 0.8351\newline0.2126\newline0.3389 & 0.8259\newline0.2176\newline0.3444 \\
     \hline
	 a=4                            & 0.663\newline0.156\newline0.2526   & 0.7366\newline0.2526\newline0.3762 & 0.7902\newline0.2366\newline0.3642 & 0.8647\newline0.1894\newline0.3108 & 0.8559\newline0.2432\newline0.3787 & 0.858\newline0.1878\newline0.3082  \\
     \hline
	 a=5                            & 0.6548\newline0.1765\newline0.278  & 0.8286\newline0.2728\newline0.4105 & 0.8365\newline0.228\newline0.3584  & 0.8731\newline0.2248\newline0.3576 & 0.8685\newline0.2487\newline0.3866 & 0.8969\newline0.2323\newline0.369 \\
    \hline

        \hline

	\end{tabularx}

\end{table}

To compare the outcomes, we present the results obtained from the Node2Vec approach in Table \ref{tab:emb}.

For the first graph, the outcomes are highly optimistic and tend to improve, particularly for larger values of $a$ and $b$, whereby the latter appears to have a more substantial influence. For nodes picked at random, we present the particulars regarding precision and recall in Table \ref{tab:gr1-rand}, and for betweennes and degree centrality in Table \ref{tab:gr3-cent}. Overall, the proposed $a,b$-model demonstrates similar precision performance compared to Node2Vec and better recall, as shown in Table \ref{tab:emb}. The values for betweenness centrality are slightly worse, while those for degree centrality are comparable. Nonetheless, these differences do not significantly affect the model.

\begin{figure}[t]
\centering
\includegraphics[width=0.45\textwidth]{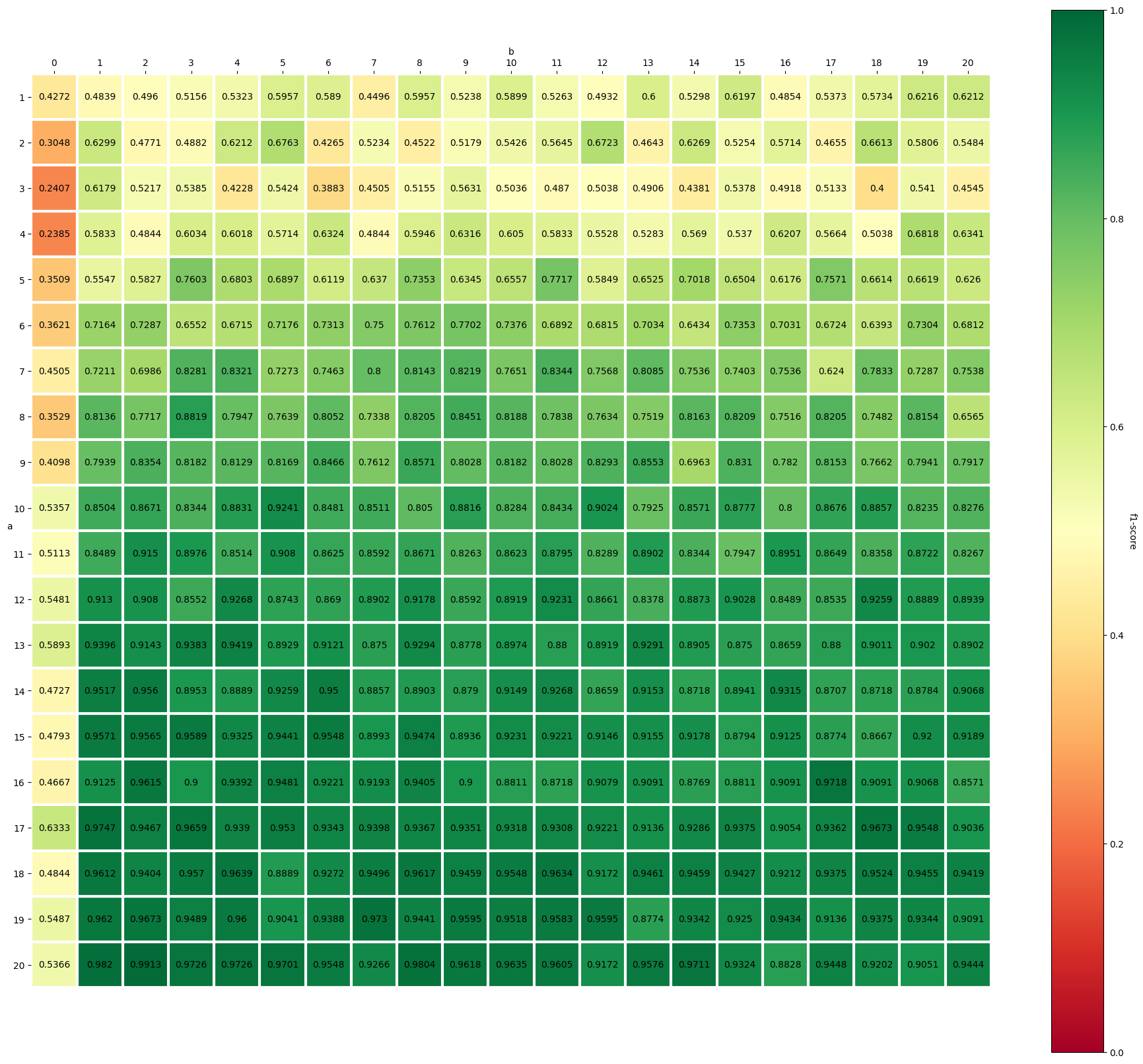}
\caption{$F_1$-score heatmap for $a,b\leq 20$ and graph 2 and randomly chosen nodes.}\label{img:resla1}

\vspace*{0.5cm}

\includegraphics[width=0.45\textwidth]{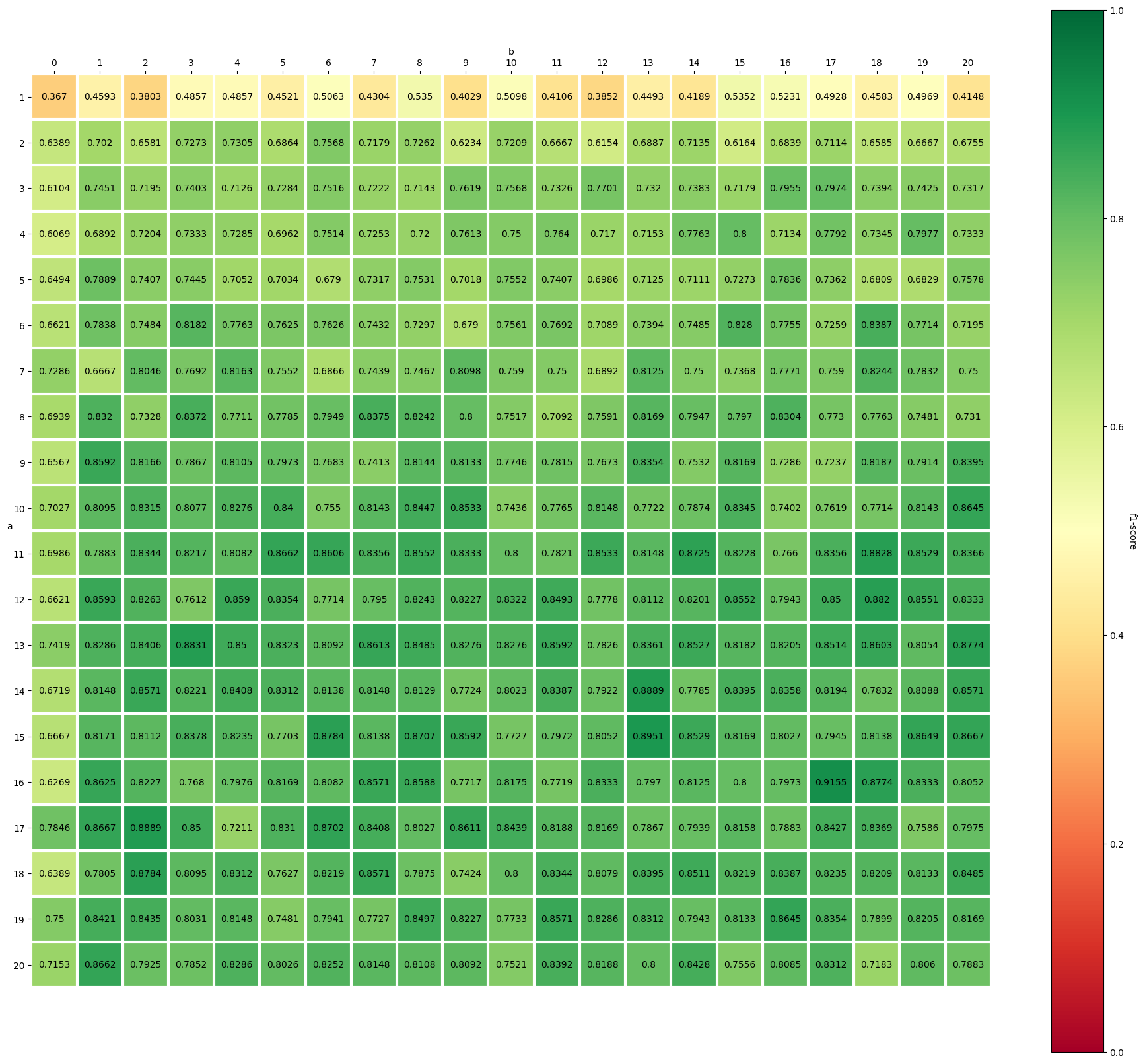}
\caption{$F_1$-score heatmap for $a,b\leq 20$ and graph 2 and degree centrality.}\label{img:resla2}
\end{figure}

The situation changes when considering Graph 2. The $F_1$-score score for Node2Vec is 0.78, as shown in Table  \ref{tab:emb}. However, Figures  \ref{img:resgr2a}-\ref{img:resgr2c} indicate the need for careful selection of values for parameters a and b to improve performance. In general, randomly selecting nodes leads to inferior performance, as demonstrated in Table  \ref{tab:gr2-rand}. The best results are achieved with betweenness centrality, as reported in Table  \ref{tab:gr2-cent}. The $a,b$-model displays superior recall value. Nonetheless, results for $a,b\leq 20$ and randomly selected nodes and degree centrality for Graph 2 are shown in Figures \ref{img:resla1} and \ref{img:resla2}. Both figures suggest that an increase in $a$ has a greater impact on $F_1$-score, but this relationship is generally less clear for degree centrality.

The situation worsens when Graph 3 is considered, as shown in Figures \ref{img:resgr3a}-\ref{img:resgr3c}. The strategy of selecting nodes at random, as shown in Table \ref{tab:gr3-rand}, proves inferior to both degree and betweenness centrality, as presented in Table \ref{tab:gr3-cent}. Although Node2Vec yields a higher $F_1$-score, the $a,b$-model incorporating centrality measures can offer better recall values.

For Graph 3, we computed higher values for parameter $a$ and $b$ utilizing degree centrality and randomly selected nodes. Specifically, by setting $a=65$ and $b=60$, we achieved a precision of 0.81 and 0.95, a recall of 0.41 and 0.67, and an $F_1$-score of 0.54 and 0.78. These results show a significant improvement over Node2Vec. Therefore, using higher values for $a$ and $b$ clearly enhances the performance.

Testing on various networks has demonstrated that the performance of the $a,b$-model relies mostly on the available data. As the mean degree of the graph grows, the performance degrades notably for lower $a$ and $b$ values. The $a,b$-model functions adeptly on graphs showcasing low density, with comparable performance to embedding. Nevertheless, embeddings outdo the $a,b$-model for small $a$ and $b$ values on dense graphs.

Using centrality measures for neighbor selection significantly boosts recall. When comparing the two measures, it is noted that degree centrality leads to better performance. Selection based on degree centrality targets nodes with the most neighbors, making it able to find many neighbors even with large $a$ and $b$. Nevertheless, as the slope of $a$ and $b$ rises, the impact of centrality measures declines. This could be because nodes with large degree provide a larger neighborhood in turn.

For all datasets, the notable positive impact on performance by increasing the parameters $a$ and $b$ is remarkable. The heatmaps of all models illustrate that the ranges $a \in [3,4]$ and $b \in [1,2,3]$ mostly exhibit the best results for $a\in [1,\dots ,5], b\in [0,\dots ,5]$. Graphs 2 and 3 demonstrate that improving the number of neighbors leads to better results.

In general, the selection of appropriate values for $a$ and $b$ presents a challenge, particularly when utilizing randomly chosen nodes or degree centrality. Additionally, it remains unclear how these centrality measures influence the technique, and further investigation should prioritize examining other centrality measures. Furthermore, assessment must encompass not just precision, recall, and $F_1$-score but also generalizability. Namely, can we extend a model that is trained on one social network to a comparable social network?

\begin{table*}[t]
\centering
\caption{Precision, Recall and $F_1$-score  for $a,b\leq 5$ and graph 1 and betweennes centrality and degree centrality.}\label{tab:gr1-cent}
	\begin{tabularx}{\linewidth}{|l|XXXXXX|XXXXXX|}
    \hline
     & \multicolumn{6}{>{\hsize=\dimexpr6\hsize+4\tabcolsep+2\arrayrulewidth\relax}X|}{Betweennes centrality} & \multicolumn{6}{>{\hsize=\dimexpr6\hsize+4\tabcolsep+2\arrayrulewidth\relax}X|}{Degree centrality} \\
    \hline
		 \, & b=0                      & b=1                      & b=2                      & b=3                      & b=4                      & b=5                      & b=0                      & b=1                      & b=2                      & b=3                      & b=4                      & b=5                      \\
        \hline

        a=1                    & 0.8772\newline0.7042\newline0.7812 & 0.8103\newline0.6912\newline0.746  & 0.8169\newline0.6905\newline0.7484 & 0.7692\newline0.6944\newline0.7299 & 0.8596\newline0.7206\newline0.784  & 0.8525\newline0.65\newline0.7376   & 0.8772\newline0.7042\newline0.7812 & 0.8704\newline0.6912\newline0.7705 & 0.8333\newline0.7143\newline0.7692 & 0.7761\newline0.7222\newline0.7482 & 0.8571\newline0.7059\newline0.7742 & 0.8525\newline0.65\newline0.7376   \\
        \hline
	 a=2                    & 1.0\newline0.8049\newline0.8919    & 0.9355\newline0.8657\newline0.8992 & 0.96\newline0.809\newline0.878     & 0.9194\newline0.9048\newline0.912  & 0.85\newline0.8831\newline0.8662   & 0.8939\newline0.9219\newline0.9077 & 1.0\newline0.7805\newline0.8767    & 0.8906\newline0.8507\newline0.8702 & 0.973\newline0.809\newline0.8834   & 0.8769\newline0.9048\newline0.8906 & 0.8701\newline0.8701\newline0.8701 & 0.8939\newline0.9219\newline0.9077 \\
     \hline
	 a=3                    & 0.9273\newline0.8793\newline0.9027 & 0.9804\newline0.8065\newline0.885  & 0.9394\newline0.9118\newline0.9254 & 0.957\newline0.978\newline0.9674   & 0.9643\newline0.8852\newline0.9231 & 0.9412\newline0.9275\newline0.9343 & 0.95\newline0.9828\newline0.9661   & 0.9455\newline0.8387\newline0.8889 & 0.9394\newline0.9118\newline0.9254 & 0.9468\newline0.978\newline0.9622  & 0.9649\newline0.9016\newline0.9322 & 0.9559\newline0.942\newline0.9489  \\
     \hline
	 a=4                    & 0.9706\newline0.9041\newline0.9362 & 0.9857\newline0.9324\newline0.9583 & 0.9853\newline0.9306\newline0.9571 & 0.9825\newline0.9333\newline0.9573 & 0.9459\newline0.9333\newline0.9396 & 0.9506\newline0.9872\newline0.9686 & 0.9559\newline0.8904\newline0.922  & 0.9857\newline0.9324\newline0.9583 & 0.9857\newline0.9583\newline0.9718 & 0.9825\newline0.9333\newline0.9573 & 0.9863\newline0.96\newline0.973    & 0.95\newline0.9744\newline0.962    \\
     \hline
	 a=5                    & 0.9231\newline0.9524\newline0.9375 & 0.9571\newline0.9178\newline0.9371 & 1.0\newline0.8871\newline0.9402    & 0.9859\newline0.9211\newline0.9524 & 0.9565\newline0.9429\newline0.9496 & 0.9552\newline0.9552\newline0.9552 & 0.9219\newline0.9365\newline0.9291 & 0.9577\newline0.9315\newline0.9444 & 0.9836\newline0.9677\newline0.9756 & 0.9853\newline0.8816\newline0.9306 & 0.971\newline0.9571\newline0.964   & 0.9688\newline0.9254\newline0.9466 \\
     \hline
	\end{tabularx}

    \vspace*{0.6cm}
    \caption{Precision, Recall and $F_1$-score  for $a,b\leq 5$ and graph 2 and betweennes centrality and degree centrality.}\label{tab:gr2-cent}
	\begin{tabularx}{\linewidth}{|l|XXXXXX|XXXXXX|}
    \hline
     & \multicolumn{6}{>{\hsize=\dimexpr6\hsize+4\tabcolsep+2\arrayrulewidth\relax}X|}{Betweennes centrality} & \multicolumn{6}{>{\hsize=\dimexpr6\hsize+4\tabcolsep+2\arrayrulewidth\relax}X|}{Degree centrality} \\
    \hline
		 \, & b=0                      & b=1                      & b=2                      & b=3                      & b=4                      & b=5                      & b=0                      & b=1                      & b=2                      & b=3                      & b=4                      & b=5                      \\
        \hline
       a=1                    & 0.4889\newline0.3235\newline0.3894 & 0.5574\newline0.4359\newline0.4892 & 0.4483\newline0.3133\newline0.3688 & 0.5352\newline0.5\newline0.517     & 0.4776\newline0.4384\newline0.4571 & 0.5424\newline0.381\newline0.4476  & 0.4878\newline0.2941\newline0.367  & 0.5439\newline0.3974\newline0.4593 & 0.4576\newline0.3253\newline0.3803 & 0.5312\newline0.4474\newline0.4857 & 0.5075\newline0.4658\newline0.4857 & 0.5323\newline0.3929\newline0.4521 \\
			\hline
             a=2                    & 0.7432\newline0.6707\newline0.7051 & 0.7536\newline0.6341\newline0.6887 & 0.7297\newline0.6667\newline0.6968 & 0.6562\newline0.5185\newline0.5793 & 0.7581\newline0.5663\newline0.6483 & 0.6795\newline0.7361\newline0.7067 & 0.6629\newline0.7195\newline0.6901 & 0.7391\newline0.622\newline0.6755  & 0.7375\newline0.7284\newline0.7329 & 0.6962\newline0.679\newline0.6875  & 0.7867\newline0.7108\newline0.7468 & 0.6552\newline0.7917\newline0.717  \\
			\hline
            a=3                    & 0.8627\newline0.4632\newline0.6027 & 0.831\newline0.7763\newline0.8027  & 0.7945\newline0.6374\newline0.7073 & 0.8406\newline0.7073\newline0.7682 & 0.7818\newline0.7414\newline0.7611 & 0.7123\newline0.6842\newline0.698  & 0.8824\newline0.4737\newline0.6164 & 0.8125\newline0.6842\newline0.7429 & 0.825\newline0.7253\newline0.7719  & 0.7763\newline0.7195\newline0.7468 & 0.7333\newline0.7586\newline0.7458 & 0.7013\newline0.7105\newline0.7059 \\
			\hline
            a=4                    & 0.8095\newline0.6456\newline0.7183 & 0.8261\newline0.6404\newline0.7215 & 0.8125\newline0.6753\newline0.7376 & 0.75\newline0.6456\newline0.6939   & 0.8413\newline0.6795\newline0.7518 & 0.6875\newline0.7143\newline0.7006 & 0.7925\newline0.5316\newline0.6364 & 0.7887\newline0.6292\newline0.7    & 0.7467\newline0.7273\newline0.7368 & 0.7436\newline0.7342\newline0.7389 & 0.8116\newline0.7179\newline0.7619 & 0.642\newline0.6753\newline0.6582  \\
			\hline
            a=5                    & 0.7679\newline0.5119\newline0.6143 & 0.831\newline0.6782\newline0.7468  & 0.8077\newline0.75\newline0.7778   & 0.8434\newline0.7292\newline0.7821 & 0.8226\newline0.6711\newline0.7391 & 0.7945\newline0.6744\newline0.7296 & 0.803\newline0.631\newline0.7067   & 0.8533\newline0.7356\newline0.7901 & 0.8358\newline0.6667\newline0.7417 & 0.7742\newline0.75\newline0.7619   & 0.7727\newline0.6711\newline0.7183 & 0.7875\newline0.7326\newline0.759  \\
     \hline
	\end{tabularx}

    \vspace*{0.6cm}
    \caption{Precision, Recall and $F_1$-score  for $a,b\leq 5$ and graph 3 and betweennes centrality and degree centrality.}\label{tab:gr3-cent}
	\begin{tabularx}{\linewidth}{|l|XXXXXX|XXXXXX|}
    \hline
     & \multicolumn{6}{>{\hsize=\dimexpr6\hsize+4\tabcolsep+2\arrayrulewidth\relax}X|}{Betweennes centrality} & \multicolumn{6}{>{\hsize=\dimexpr6\hsize+4\tabcolsep+2\arrayrulewidth\relax}X|}{Degree centrality} \\
    \hline
		 \, & b=0                      & b=1                      & b=2                      & b=3                      & b=4                      & b=5                      & b=0                      & b=1                      & b=2                      & b=3                      & b=4                      & b=5                      \\
        \hline
        a=1                    & 0.5289\newline0.2587\newline0.3475 & 0.5398\newline0.2979\newline0.3839 & 0.5158\newline0.3113\newline0.3883 & 0.5\newline0.3118\newline0.3841    & 0.5127\newline0.3389\newline0.4081 & 0.503\newline0.346\newline0.41     & 0.5089\newline0.2934\newline0.3722 & 0.5092\newline0.3324\newline0.4023 & 0.5491\newline0.3482\newline0.4261 & 0.494\newline0.325\newline0.3921   & 0.4991\newline0.37\newline0.425    & 0.5039\newline0.3514\newline0.4141 \\
		\hline
        a=2                    & 0.5431\newline0.2869\newline0.3754 & 0.5795\newline0.336\newline0.4254  & 0.5438\newline0.3518\newline0.4272 & 0.5373\newline0.3553\newline0.4277 & 0.4957\newline0.387\newline0.4347  & 0.5418\newline0.4018\newline0.4614 & 0.5535\newline0.3673\newline0.4416 & 0.5756\newline0.386\newline0.4621  & 0.5369\newline0.4034\newline0.4606 & 0.5256\newline0.4088\newline0.4599 & 0.5189\newline0.4016\newline0.4528 & 0.5587\newline0.4367\newline0.4902 \\
		\hline
        a=3                    & 0.6113\newline0.2747\newline0.379  & 0.6111\newline0.3419\newline0.4385 & 0.631\newline0.364\newline0.4617   & 0.572\newline0.3595\newline0.4415  & 0.5669\newline0.3727\newline0.4497 & 0.5758\newline0.3735\newline0.4531 & 0.5597\newline0.356\newline0.4352  & 0.5594\newline0.4135\newline0.4755 & 0.5363\newline0.4093\newline0.4643 & 0.5204\newline0.3915\newline0.4468 & 0.5215\newline0.4291\newline0.4708 & 0.5078\newline0.4246\newline0.4625 \\
		\hline
        a=4                    & 0.7051\newline0.2861\newline0.407  & 0.6755\newline0.3652\newline0.4741 & 0.6509\newline0.3316\newline0.4393 & 0.6759\newline0.3466\newline0.4583 & 0.6474\newline0.3759\newline0.4757 & 0.6635\newline0.3507\newline0.4589 & 0.6056\newline0.3355\newline0.4318 & 0.6004\newline0.4385\newline0.5068 & 0.5593\newline0.4037\newline0.4689 & 0.5576\newline0.3995\newline0.4655 & 0.5572\newline0.4169\newline0.4769 & 0.5643\newline0.4204\newline0.4818 \\
		\hline
        a=5                    & 0.6997\newline0.2996\newline0.4195 & 0.6787\newline0.3548\newline0.466  & 0.6694\newline0.3263\newline0.4388 & 0.6792\newline0.3294\newline0.4437 & 0.687\newline0.3481\newline0.4621  & 0.6838\newline0.3378\newline0.4522 & 0.6521\newline0.3666\newline0.4694 & 0.5943\newline0.4234\newline0.4945 & 0.5529\newline0.3906\newline0.4578 & 0.5642\newline0.3791\newline0.4535 & 0.5578\newline0.4086\newline0.4717 & 0.5694\newline0.4219\newline0.4847  \\

     \hline
	\end{tabularx}
                   \end{table*}

\section{Discussion and outlook}
The study suggests a promising approach using neighboring nodes as features in the feature vector in various forms, with and without centrality measures. Comparing the embeddings on different data, supports the effectiveness of the a,b-model. It generally outperforms embeddings for large values of a,b. However, in some aspects, such as recall on sparse graphs, it still generally outperforms embeddings.

The research question initially aimed to explore this topic. How can we combine depth and breadth search methods to identify the context, which is the structure of node neighborhoods, in order to enhance machine learning techniques for predicting edges in graphs? We introduced a promising model that utilizes the neighbors of nodes as characteristics in the feature vector with and without the use of centrality measures in various forms. Through comparison with embeddings from different datasets, we confirmed that the $a,b$-model demonstrates promising results. It appears that it generally outperforms embeddings for high values of $a,b$. In terms of \textit{recall} on sparsely connected graphs, it also seems to outperform embeddings even for low values.

However, the optimal node selection strategy for a node's neighborhood remains unclear. Additionally, the number of neighbors to consider remains unclear. Although we have demonstrated significant variation in outcomes using different methods, random node selection and degree centrality show promising results. Nevertheless, we have not been able to prove their generalizability.  Generally, larger values of $a$ and $b$ tend to provide better outcomes; however, this is highly reliant on the underlying graph. Whereas smaller values yield favorable results for graphs with a small average node degree, it is imperative to increase these values considerably.

Therefore, the model's practical application depends on the data structure and task. Thinner graph tables demonstrate that utilizing centrality measures leads to a decline in precision for \textit{Recall} slope.  This means that while more links are predicted, the likelihood of accurately predicting them decreases. Models utilizing centrality measures can predict less sensitive data, such as in streaming service recommendation algorithms or social network friend suggestions. This expands the range of recommendations for users. However, inaccuracies in these suggestions are not critical. In contrast, accuracy is crucial in medical predictions. Diseases are less likely to be predicted solely based on symptoms, however, predictions obtained are more likely to be accurate. In thinner graphs, embeddings are more dominant compared to $a,b$ models. They exhibit higher precision and only slightly lower recall than the $a,b$ model with centrality measures.

Although our proof of concept is both functional and generic, further research on this approach is necessary.

\bibliography{lit}

\end{document}